# Teleconnection processes linking the intensity of the Atlantic Multidecadal Variability to the climate impacts over Europe in boreal winter

Saïd Qasmi, Christophe Cassou and Julien Boé
*CECI, Université de Toulouse, CNRS, Cerfacs, Toulouse, France*

*Corresponding author address:* Saïd Qasmi, Centre National de Recherches Météorologiques, 42 Avenue Gaspard Coriolis, 31057 Toulouse, France.
E-mail: qasmi@cerfacs.fr

# ABSTRACT

The teleconnection between European climate and Atlantic Multidecadal Variability (AMV) remains difficult to isolate in observations because of internal variability and anthropogenically-forced signals. Using model sensitivity experiments proposed within the CMIP6/DCPP-C framework, the wintertime AMV/Europe teleconnection is investigated in large ensembles of pacemaker-type simulations in the CNRM-CM5 global circulation model. To evaluate the sensitivity of the model response to the AMV amplitude, experiments with AMV-forcing pattern multiplied by 2 and 3 (hereafter 2xAMV and 3xAMV, respectively) are performed in complement to the reference ensemble (1xAMV). Based on a flow analog method, the AMV-forced atmospheric circulation is found to cool down the European continent, whereas the residual signal, mostly including thermodynamical processes, contributes to warming. In 1xAMV, both terms cancel each other, explaining the overall weak AMV-forced atmospheric signal. In 2xAMV and 3xAMV, the thermodynamical contribution overcomes the dynamical cooling and is responsible for milder and wetter conditions. The thermodynamical term includes the advection of warmer and more humid oceanic air penetrating inland and the modification of surface radiative fluxes linked to (i) altered cloudiness and (ii) snow-cover reduction acting as a positive feedback with the AMV amplitude. The dynamical anomalous circulation combines (i) a remote response to enhanced diabatic heating acting as a Rossby-wave source in the western tropical Atlantic and (ii) a local response associated with warmer SST over the subpolar gyre favorizing an anomalous High. The weight between the tropical-extratropical processes and associated feedbacks is speculated to partly explain the nonlinear sensitivity of the response to the AMV-forcing amplitude, challenging thus the use of the so-called pattern-scaling technique.

# 1. Introduction

The importance for many societal applications of improved information about near-term climate evolution (from 1 year to a decade in advance) has prompted considerable research in the field of decadal climate prediction (e.g. Meehl et al. 2014). In the framework of the fifth Coupled Model Intercomparison Project (CMIP5, Taylor et al. 2012), decadal forecast experiments initialized with observationally based state information have shown greater skill in predicting the evolution of planetary-averaged temperature compared to traditional non-initialized historical coupled simulations (Kirtman et al. 2013, Bellucci et al. 2015). Beyond global and integrative metrics, robust skill in hindcasts is also found at ocean/continental basin-scale for leadtimes up to 7-8 years and for particular regions such as the North Atlantic sector, which clearly stands out (Doblas-Reyes et al. 2013). On top of prescribed anthropogenic and natural external forcings (Terray 2012), the so-called Atlantic Multidecadal Variability (AMV), characterized by basin-wide low-frequency variations of the North Atlantic sea surface temperature (SST), has been identified as one of the sources of predictability at decadal timescale (e.g. Kim et al. 2012).

The origin of the AMV is still highly debated due to the shortness, spatial sparsity and uneven quality of the observational record over the instrumental epoch (e.g. Cassou et al. 2018) and additionally, to the probable coexistence and combination of several physical mechanisms yielding low-frequency fluctuations over the North Atlantic, as assessed from modeling results (see Yeager and Robson 2017 and Zhang et al. 2019 for a review). Although the integration of so-called atmospheric noise by the ocean has been recently proposed as a potential source of decadal variability in the North Atlantic (Clement et al. 2015; Cane et al. 2017), part of the observed AMV is commonly considered as the surface fingerprint of ocean heat content anomalies driven by internal climate dynamics (O'Reilly et al. 2016). This involves large-scale changes in both air-sea fluxes and ocean heat transport through the variability of the North Atlantic subpolar and subtropical horizontal gyres and the Atlantic Meridional Overturning Circulation (AMOC) (e.g. Zhang and Wang, 2013) in presence of complex feedbacks (Ruprich-Robert and Cassou, 2015, Peings et al. 2016). Consistently, Msadek et al. (2014), Robson et al. (2012), Yeager et al. (2015) among others, show that the prediction of the North Atlantic SST and ocean heat content as well as sea ice extent in subarctic basins clearly benefits from the initialization of the 3-dimensional thermodynamical ocean.

However, the added value of the initialization is considerably reduced over the North Atlantic adjacent continents, as found in most of CMIP5 decadal prediction systems (Goddard et al. 2013, Doblas-Reyes et al. 2013). Such a loss of predictability over land is somewhat paradoxical given the tight links that exist in observations between AMV and the decadal variations in summertime temperature and precipitation over the North American continent (Sutton and Hodson 2005, Ruprich-Robert et al. 2017), over Europe (Sutton and Dong 2012, O'Reilly et al. 2017) and over Africa for Sahel rainfall (Zhang and Delworth 2006). Note that greater predictive model performance is found for specific decadal shifts [e.g. the mid-1990 warming of the subpolar gyre (Robson et al. 2013)], but despite the existence of such a conditional skill, the use of decadal prediction systems yet remains limited for operational purposes (Towler et al. 2018).

More optimistic views and opportunities for progress in the science of decadal forecast have been recently presented in Yeager et al. (2018). Based on an updated version of the CESM model, they insist on the importance of minimizing intrinsic model biases and inhomogeneities in initialized fields in key regions, such as the North Atlantic Ocean, in order to limit spurious drifts and shocks (Sanchez et al. 2016), which deteriorate the levels of skill. They also clearly illustrate the crucial need for large ensembles to robustly extract the predictive signals over land that can be attributable to the initialized decadal ocean fluctuations. Despite these new promising results, a key outstanding challenge for the climate research community is to better understand how decadal changes in the ocean affect surface climate over land and ultimately translate into useful prediction. Over the North Atlantic, obstacles stand in the diversity of the statistical (amplitude, intrinsic frequency, etc.) and physical (spatial patterns, role of AMOC, etc.) properties of the AMV simulated by the current generation of models, leading to large uncertainties in teleconnectivity over Europe as shown in Qasmi et al. (2017). In their study based on long control model simulations and historical ensembles, they further insist on the non-stationarity of the level, even sign, of the AMV-Europe teleconnection as a function of the considered period; this clearly adds a degree of complexity to extract the AMV-forced fingerprint and associated physical processes at the origin of the observed ocean-land relationship.

As above-listed, most of the robust AMV impacts over Europe have been reported for summertime but the seasonality of the teleconnection remains an open question. Based on observations, contradictory results may be found in literature for winter. For instance, Sutton and Dong (2012) could not find any significant anomalous atmospheric circulation over the North Atlantic in concomitance with AMV phases and they claim that no significant signal in temperature and precipitation could be detected over Europe. O'Reilly et al. (2017) confirm the missing continental AMV-fingerprint and attribute the lack of teleconnectivity to the dominance of atmospheric noise whose intensity/weight is maximum in winter and may thus overcome any potential oceanic influences. Based on a lagrangian approach, Yamamoto and Palter (2016) alternatively interpret the "seasonal teleconnectivity hole" as the result of compensation between AMV-driven anomalies in atmospheric dynamics on the one hand and direct thermodynamic influence through air-sea fluxes on the other hand. Introducing temporal lags between atmospheric and oceanic fields in the observations, Gastineau and Frankignoul (2015) suggest that the large-scale wintertime atmosphere response to positive AMV (hereafter AMV+) projects onto the negative phase of the North Atlantic Oscillation (NAO-). Peings et al. (2016) find a similar response but only for two models spanning the full CMIP5 archive. They attribute the weak feedback of the AMV onto the wintertime atmosphere to the coupled model deficiencies in generating strong-enough and persistent-enough multidecadal variability over the North Atlantic in line with Kavvada et al. (2013), Qasmi et al. (2017), among others.

Results appear more robust in dedicated sensitivity model experiments with prescribed or restored SSTs. Peings and Magnusdottir (2014) provide evidence for favoured NAO- (NAO+) circulation regimes during AMV+ (AMV-) and Davini et al. (2015), consistently with earlier studies (e.g. Cassou et al. 2004, Hodson et al. 2011), interpret this relationship as a by-product of forced atmospheric Rossby waves generated in the Caribbean basin by altered convection in response to AMV-related SSTs anomalies. Peings et al. (2016), similarly to Drévillon et al. (2003), confirm the importance of ocean-atmosphere feedbacks at midlatitudes to allow a full northward extension of the tropical-initiated Rossby wave in order to generate significant impacts over Europe located at the tail end of the teleconnection. The relative importance/weight of tropical versus extratropical AMV-related SST anomalies is also analysed in Ruprich-Robert et al. (2017) but the latter study clearly insists on the overall weak signal-to-noise (SNR) ratio in terms of wintertime atmospheric response and *a fortiori* AMV-forced teleconnectivity over Europe. Being a true a.k.a. intrinsic characteristic of the climate system, or a deficiency of the current generation of models, that "teleconnectivity hole" remains a key scientific question, which definitely conditions the level of potential predictability in decadal forecast systems as raised in Yeager et al. (2018).

Within this context and built on lessons drawn from CMIP5, the Decadal Climate Prediction Project (DCPP) has proposed for CMIP6 (Eyring et al. 2016) a new targeted multi-model framework (named Component C, Boer et al. 2016) aiming at increasing knowledge and physical understanding of the worldwide impacts of the AMV through teleconnectivity (Cassou et al. 2018). The CMIP6-endorsed coordinated experiments are inspired from Ruprich-Robert et al (2017) and rely on so-called pacemaker simulations where the North Atlantic SSTs are restored towards a specific anomalous pattern that is representative of AMV phases, whereas the rest of the coupled model remains free to evolve. We here conducted those specific DCPP-Component-C experiments using the CNRM-CM5 global circulation model (Voldoire et al. 2013). In the following paper, we concentrate our analyses and physical interpretations of the model AMV-forced response in winter over Europe. We employ a so-called "circulation analog technique" inspired from Boé et al. (2009), Deser et al. (2016) and O'Reilley et al (2017) to decompose the impact of the AMV on surface air temperature and precipitation over Europe into dynamical versus so-called thermodynamical relative contributions. Considering the weak SNR properties documented in many studies, we have performed additional experiments in which the AMV-related SST anomalies are artificially boosted to potentially increase the forced response. Concurrently, we have produced larger ensembles than current protocols recommend for CMIP6 in order to get a better estimation of the AMV-forced response in presence of prevalent climate noise, following Deser et al. (2012) and Yeager et al. (2018) advices.

The structure of the paper is organized around three main objectives: (i) isolate the dynamical and thermodynamical fingerprints of the AMV in the North Atlantic/European climate assessed from our large ensembles and revisit the results presented in Ruprich-Robert et al. (2017), Yamamoto and Palter (2016) and O'Reilly et al. (2017), (ii) identify the physical processes explaining the modelled AMV teleconnectivity over Europe in winter, (iii) evaluate the sensitivity of the model atmospheric response and related mechanisms to the amplitude of the AMV. After a description of the modelling protocols in section 2, the mean wintertime response to the AMV, as well as the dynamical and thermodynamical mechanisms, are detailed in section 3. The sensitivity of the AMV-forced atmospheric response to the AMV amplitude is discussed, followed by conclusions and perspectives in section 4.

## 2. Methods

*a. Model pacemaker sensitivity experiments*

As mentioned in the Introduction, the pacemaker or partial coupling simulations analyzed in the paper follows the protocol endorsed by CMIP6 and commonly labelled as DCPP-C AMV experiments. The reader is invited to refer to Boer et al. (2016, Components C1.2 and C1.3 in their Table C1) for a throughout presentation of the coordinated experimental framework and related input datasets shared through input4MIPs (Durack et al. 2018). Fig. 1a shows the AMV SST anomalous fingerprint towards which the models are restored over the North Atlantic. In CNRM-CM5, this is achieved through the addition of a feedback term to the non-solar total heat flux in the surface temperature derivative equation following Haney (1971). This flux formulation affects the entire ocean mixed layer depth. In compliance with DCPP recommendations (see technical notes in https://www.wcrp-climate.org/experimental-protocol), we set the restoring coefficient to a spatially and temporally constant value equal to -40 W m$^{-2}$ K$^{-1}$. For the SST, it is equivalent to a damping time scale of ~2 months for a 50-m deep mixed layer.

Two large ensembles of 40 members of 10-year long simulations are conducted. They differ by the sign of the targeted anomalous SST pattern, corresponding to either positive (warm, hereafter AMV+ ensemble) or negative (cold, hereafter AMV-, i.e. sign-reversal of Fig.1a) phases of the AMV. The initial conditions for the 40 members are ocean+atmosphere+land+sea ice states (so-called macro-perturbation following Hawkins et al. 2016 nomenclature) taken arbitrarily every 20 years from the 850-yr long CMIP5 control preindustrial experiment of CNRM-CM5. The same set of initial conditions is used for the AMV+ and AMV- ensembles. The ensembles size has been increased here to 40 instead of 25 (minimum number recommended in DCPP-C, Boer et al. 2016) to ensure a better estimation of AMV-forced signals. Additional twin experiments are conducted by multiplying by 2 and 3 the anomalous SST pattern towards which the model is restored over the North Atlantic (Fig.1a). Those additional ensembles are hereafter termed 2xAMV and 3xAMV, respectively, and the reference DCPP-compliant ensemble is referred to as 1xAMV.

Fig. 1b provides a crude evaluation of the pacemaker protocol and importantly, an indication of the actual SST forcing in each ensemble. Independently of the sign of the AMV experiments, a spread exists in simulated annual SSTs for all ensembles. The corresponding ensemble means are always lower than the targeted SSTs towards which the coupled model is restored. Both features are attributable to the weak restoring coefficient used here. We tested stronger values, which do allow the actual SST to be closer to the targeted SST (not shown). However, those lead to spurious energy imbalance, perturb the modeled high frequency air-sea interactions, the ocean heat content and meridional transports through AMOC etc., which ultimately affect the interpretation of the atmospheric response to the AMV forcing (see also DCPP technical notes, https://www.wcrp-climate.org/wgsip/documents/Tech-Note-2.pdf). Despite a weak restoring coefficient, the interannual variance of the modeled SST averaged over the North Atlantic is reduced by a factor of 10 compared to the free piControl CNRM-CM5 experiment. Because the restoring coefficient is fixed, its efficiency is strongly dependent on the ocean mixed layer depth. The reduction of the basin-wide variance thus masks considerable regional heterogeneities between, for instance, the subpolar gyre characterized by seasonal deep ocean mixing and the more stratified tropical Atlantic regions (see also Ruprich-Robert et al. 2017, Ortega et al. 2017).

Since the restoring is not perfect, the multiplication by 2 or 3 of the anomalous SST-forcing pattern in our additional ensembles is not as artificial or unrealistic as it may appear at first glimpse. Actual North Atlantic SSTs obtained in 3xAMV are in fact close to the targeted SSTs of 2xAMV, which correspond to +/- two standard deviations of the observed AMV index over the instrumental record. Actual SSTs in 2xAMV are close to the targeted SSTs of 1xAMV (Fig. 1b). These additional experiments will be useful to assess the sensitivity of the teleconnection to the intensity of the AMV-forced SST anomalies and in particular, its degree of linearity.

*b. Flow analog technique*

Boé et al. (2009), Cattiaux et al. (2013), Deser et al. (2016), among others, employed so-called "dynamical flow analogs" techniques to quantify the relative roles of the dynamical versus non-dynamical processes in either observed or projected climate change signals. More recently, O'Reilly et al. (2017) apply the same technique to study the AMV teleconnection over North America, Europe and Africa based on observational datasets (reanalyzes) over the historical period. We here adapt the methodology to our ensemblist approach aiming at better understand the involved mechanisms of the AMV-forced surface temperature and precipitation anomalies simulated in CNRM-CM5 over Europe in winter.

Technically, for each winter day $K$ of the AMV+ experiment, we seek for the $N$ best analogs of the atmospheric circulation in the population of winter days from the twin AMV- experiment. We use Sea Level Pressure (SLP) maps centered over Europe (EU, 35°-70°N, 15W-35E) and the similarity criterion to define the circulation analogs is the Euclidean distance. The $N$ best analogs for $K$ are the $N$ days in AMV- for which the Euclidean distances to $K$ are minimum. We then reconstruct the temperature/precipitation map of Day $K$ of AMV+ by averaging the $N$ temperature/precipitation maps of the best $N$ SLP analogs found in AMV-. Assuming the absence of feedback processes between the surface and the circulation, the latter reconstructed temperature/precipitation is interpreted as the surface fingerprint of the atmospheric circulation (hereafter named the dynamical part of the field) and the residual with respect to the actual raw AMV+ temperature/precipitation is treated as the thermodynamical part for sake of simplicity. Note that to account for the seasonal cycle of the reconstructed surface fields, which could be particularly pronounced (e.g. for temperature), the analog search for AMV+ day $K$ is constrained to be in a 15-day window around day $K$ in the AMV- pool of days as done in Dayon et al. (2015) for instance. To sum up, let us take a concrete example. Let Day $K$ be Feb. 1$^{st}$ of Year 4 of Member 18 (01Feb-y4-m18) of AMV+. Let $N=2$. Research of analogs is done in the pool of days formed by the 40 members and 10 years of AMV- between the 24$^{th}$ of January and the 7$^{th}$ of February. We find the two best SLP analogs be 29Jan-y2-m38 and 06Feb-y10-m3 and average the corresponding raw temperature/precipitation of those 2 days of AMV- to form the reconstructed dynamical temperature/precipitation of day $K$ of AMV+. The computation is repeated for all the winter (Dec. 1$^{st}$ to Feb. 28$^{th}$) days of AMV+.

Sensitivity tests to (i) the spatial domain used for analog seek and (ii) the number $N$ of retained analogs used for reconstruction have been conducted *a priori*. To do so, the above-described stepwise process is applied within the AMV+ ensemble itself; this additionally provides a quantitative evaluation/validation of the proposed methodology. Technically speaking, a given day $K$-y$Y$-m$M$ of AMV+ is here reconstructed from the $N$ best SLP analogs found in the pool of AMV+ days excluding in that case the year $Y$ to which day K belongs to account for the day-to-day persistence of the circulation. Regarding the geographical domains, results from SLP analog extracted from a larger region corresponding to the so-called North Atlantic-Europe region (NAE, 20°-80°N, 80°W-30°E) used traditionally for large-scale dynamical purposes (see for instance, Cassou et al. 2011, Michel and Riviere 2014, etc.) have been contrasted to the above-mentioned EU sector used as reference.

A 2-step evaluation of the performance of the methodology is carried out based on spatial root mean square error (RMSE) and spatial correlation metrics between (i) the reconstructed SLP with the analog method (the predictor) and the actual SLP in AMV+ (Table 1) and (ii) the reconstructed T2m (the predictand) with the actual one in AMV+ (Table 2). For SLP reconstruction, we show that the EU domain clearly outperforms the NAE one with lower RMSE and higher correlation whatever the number of selected analogs (Table 1). The optimal number of analogs $N$ lies between 10 and 15 since (i) the highest correlation value is found for $N=10$ analogs and becomes insensitive to the inclusion of additional ones and (ii) RMSE is concomitantly the smallest for $N=10$, being slightly degraded with increased number. Results appear to be much less sensitive to the geographical domain for T2m but the overall above-listed conclusions for the choice in $N$ still hold (Table 2). To further verify the robustness of the method, all these validation steps are also repeated with AMV- instead of AMV+ experiments and additionally with the 2xAMV and 3xAMV ensembles. Results remain valid whatever the case (not shown) and the combination EU domain/$N=10$ is therefore retained for our study.

In the rest of the paper, the AMV-forced anomalies for any fields (also named response for short) are defined as the ensemble mean differences between AMV+ and AMV- experiments. The dynamical component of the AMV-forced anomalies is defined as the ensemble mean difference between the reconstructed field of AMV+ based on SLP analog seek in the counterpart AMV- experiment and the reconstructed field of AMV- based on SLP analog seek in AMV- itself. This accounts for the methodology error linked intrinsically to the analog technique. The thermodynamical component of the AMV-forced anomalies is defined as the residual anomaly calculated by subtracting the dynamical AMV-forced anomaly from the full response.

## 3. Results

*a. Mean wintertime response to AMV over Europe*

Fig. 2 summarizes the AMV-forced winter anomalies for some surface and dynamical atmospheric fields.

For surface air or 2-meter temperature (T2m), an overall weak response is obtained in CNRM-CM5 in 1xAMV (Fig. 2a) in consistence with previous studies, which highlighted the absence of detectable impact of the AMV on wintertime European climate as assessed both from models (Ruprich-Robert et al. 2017) and observations (Yamamoto and Palter 2016, O'Reilly et al. 2017). Despite very marginal point-wise significance, which is limited to a weak warming over the Atlantic side of the Iberian Peninsula, note though that a robust spatial pattern interestingly emerges at continental scale. Cooling dominates in Central Europe, from Western Russia to the North Sea shoreline including the Alpine region and South Sweden, whereas warming occurs in the northernmost part of Scandinavia and along the Mediterranean Sea, to a lower extent. Similar qualitative conclusions can be drawn in terms of precipitation (Fig. 2d). In 1xAMV, the AMV-forced response is overall weak but characterized by large-scale coherence. Drier conditions extend from the UK/northern France to Sweden/the Baltic shore of Finland where the AMV-forced response is the most pronounced. This contrasts to wetter winters along the western windward coast of Scandinavia and around the entire Mediterranean Sea with regional features that are indicative of orographic effects.

Consistently with colder and wetter conditions, albeit weak and insignificant, increased snow cover is found on the south side of the Baltic Sea along a narrow latitudinal band from the Netherlands to Belarus (Fig. 2g). Over the ocean, sea-ice extent is diminished in all the Nordic

Seas with maximum amplitude of the AMV-response along the seasonal ice edge. In terms of atmospheric circulation (Fig. 2j), higher geopotential height dominates the northern part of the Atlantic basin at 500 hPa (Z500*) with maximum loading between Iceland and the UK. Note that Z500 zonal means have been retrieved to account for the mean dilatation of the atmosphere due to the artificial heat source introduced in the model in pacemaker experiments via the flux restoring term. The signal is barotropic with a nominal eastward shift at the surface (significant higher mean sea level pressure -MSLP- centered in the North Sea), but baroclinic over the retracted sea-ice regions (Labrador and Greenland Seas). Negative MSLP and Z500* anomalies cover most of the European continent from the Iberian Peninsula to Western Russia south of 50°N.

In 2xAMV, the AMV-forced temperature anomalies are positive over the entire continent with maximum loading in Scandinavia over Sweden/Finland and along the Atlantic flank of Europe (including the entire Iberian Peninsula, Fig. 2b). Continental-scale warming is further intensified in 3xAMV (Fig. 2c) and penetrates deeper inland with significant and amplified response along an axis going from the Baltic Sea up to Southern France/Northern Spain. For precipitation, despite limited point-wise significance, wetter conditions tend to prevail over the entire continent (except Scandinavia) with maximum anomalies found over the Balkans in 2xAMV (Fig. 2e). Interestingly, although precipitation further increases on average over Europe in 3xAMV, the regional structure of the response greatly differs from the other two ensembles (Fig. 2f). Maximum excess is not found anymore in the Balkans like in 2xAMV but over Eastern Spain and a large part of Central Europe along the stretch of maximum warming (Fig. 2c).

Strong reduction of snow cover (Fig. 2i) is also collocated with the greatest positive temperature anomalies, which is indicative for enhanced rainfall at the expense of snowfall in 3xAMV. This is less valid for 2xAMV where the reinforced precipitation over the Balkans is accompanied by locally increased snow cover, yet marginally (Fig. 2h). Loss of sea ice is considerably reinforced with the amplitude of the AMV (Fig. 2hi) on both sides of the Atlantic basin, with maximum ice decline in the Odden region at the eastern edge of the Greenland Sea and along the Eastern Labrador current. In terms of atmospheric dynamics, lower Z500* and negative SLP anomalies are considerably reinforced south of 55°N and become significant from Newfoundland to the Mediterranean Sea in 2xAMV (Fig. 2k). Note though that both positive MSLP and Z500* anomalies are northwestward shifted compared to 1xAMV with a degree of intensification and significance at polar latitudes that is considerably less than their negative counterpart to the south. In 3xAMV, the AMV-forced signal in Z500* is wavier with two cyclonic cores (one between Newfoundland and the Azores and a second one over Western Russia), which sandwiches positive Z500* anomalies from Greenland to France (Fig. 2l). At the surface, stronger negative MSLP anomalies covers most of the Atlantic Ocean except in the Norwegian Sea and over Greenland where positive MSLP signals, yet slackened, remain.

To deepen our understanding of the full response in temperature and precipitation over Europe, we use in the following section the flow analog method described in section 2.b to assess the respective weight of the dynamical versus thermodynamical related processes.

*b. Decomposition in dynamical and thermodynamical components of the AMV-forced anomalies over Europe*

In 1xAMV, the atmospheric dynamical response is characterized by an anticyclonic circulation centered over Scotland, which leads to prevalent northeasterly wind anomalies over most of Europe, except Scandinavia (Fig. 3a). The latter flow is responsible for dominant, yet marginally significant, negative temperature anomalies over the entire continent. The presence of anomalous High just off Europe tends to favor the advection of colder and drier air from the East and/or to block storms to penetrate inland, which explains the large-scale deficit in rainfall along an axis going from Northern France to the Baltic shore of Sweden/Finland (Fig. 3g). Dominant northeasterlies over the Mediterranean basin lead to onshore anomalous flow over Spain leading locally to wetter significant conditions. Elsewhere, slight enhanced rainfall dominates with some orographic effect over the Carpathians and the Balkans. The thermodynamical component of the AMV-forced T2m signal is characterized by large-scale warming with maximum loading along the Atlantic shore and in the northernmost part of Scandinavia (Fig. 3d). It counteracts the dynamical component dominated by chillier conditions (Fig. 3a), leading *in fine* to a weak total response in temperature (Fig. 2a), in line with Yamamoto and Palter (2016)'s findings. For precipitation, the thermodynamical part (Fig. 3j) reinforces the dynamically-induced wetter conditions along the Mediterranean Shore (Fig. 3g). It is also responsible for increased snowfall along the South shore of the Baltic Sea (Fig. 2g), while the dynamical part clearly explains the largest portion of the drier conditions over most of Scandinavia (compare Fig. 3gj with Fig. 2d).

The storyline is very different for 2xAMV and 3xAMV temperature since the thermodynamical component controls most of the large-scale AMV-forced total warming found over the entire Europe (compare Fig. 3ef with Fig. 2bc) and thus clearly outpaces the dynamical one. The inland penetration of the thermodynamical signal is clearly function of the amplitude of the AMV SST forcing with some amplification over Central Europe, as noted earlier from Fig. 2c. Conversely, the strength of the dynamical cooling is similar in all ensembles (Fig. 3abc) and does not increase with the AMV forcing; in 3xAMV, it is even barely significant. The dynamical cooling is very much sensitive to subtle changes that occur in the position of the AMV-forced MSLP anomalies. The positive core around 60ºN progressively shifts northwestward with the amplitude of the AMV forcing, while anomalous cyclonic circulations further South move to the West from the Black Sea in 1xAMV to the Adriatic Sea in 2xAMV and off Portugal in 3xAMV. These displacements, without significant simultaneous amplification, control the strength and direction of the dominant easterly anomalies over Europe; they explain a large portion of the regional changes in the dynamical component as a function of AMV forcing.

The above conclusions for temperature are also valid for precipitation (Fig. 3hi). At large-scale, the anomalous AMV-forced circulation is responsible for wetter conditions over the European Mediterranean coast with maximum response in 2xAMV where minimum MSLP and associated cyclonic flow are the most pronounced. Concurrent drier conditions prevail north of 50ºN and are related to the anomalous advection of dry and cold air from the East or to the reduced penetration of warm and humid air masses from the West. These dynamical features are found in all ensembles, along with a southward displacement of the stormtrack over Europe (not shown). The thermodynamical response tends to increase with the amplitude of the AMV SST forcing and leads to wetter conditions at continental scale, except over the Mediterranean domain where signals are very weak (Fig. 3kl). At first glimpse, dynamical and thermodynamical contributions oppose each other in 2xAMV (Fig. 3hk) whereas clear rainfall excess (Fig. 3l) dominates in 3xAMV in Central Europe, in collocation with the area of maximum warming and snow-cover reduction extending from Catalonia to the Baltic countries, as above documented

(Fig. 2cf).

In the following sections, we concentrate on the physical mechanisms and phenomena at the origin of the dynamical and thermodynamical responses as a function of the amplitude of the AMV forcing.

*c. Mechanisms for the dynamical component of the AMV-forced response*

Zonally averaged anomalies over the North Atlantic basin are presented in Fig. 4 as function of height and as a function of the intensity of the AMV forcing. In 1xAMV, the warming imposed at the surface ocean is exported throughout the entire atmospheric column with maximum signals in the subtropics between 20ºN and 30ºN and more importantly at high latitudes from 50ºN northward (Fig. 4a). In the polar regions, there is a clear amplification of the atmospheric temperature response to the restored SST anomalies, which is caused by the pronounced AMV-forced reduction of sea ice acting as an additional source of heat at the surface in all the subarctic basins (Fig. 2g). In response to warmer SST, humidification occurs in the lower atmosphere and is exported upward to the upper troposphere between the equator and 15ºN (Fig. 4b); this is collocated with a reduction of the mean upper-level westerlies in the deep tropics (Fig. 4c). At higher latitude, albeit barely significant, the AMV-forced response is characterized by a weakening on the northern flank of the climatological maximum westerly jet around 45ºN while no signal is found elsewhere.

Amplification of the AMV-forced response is found throughout the depth of the troposphere in 2xAMV and 3xAMV for temperature (Fig. 4dg) and specific humidity (Fig. 4eh). The intensification of the signals is rather linear with respect to the amplitude of the AMV forcing. Such a linear behavior is also valid for the reduction of the westerly wind in the subtropics, whereas a different picture emerges in the extratropics (Fig. 4fi). A clear meridional dipole in zonal wind is found straggling the climatological jet core in both 2xAMV and 3xAMV, with a significant strengthening on its equatorward side and a slackening on its poleward side, implying an equatorward shift of the midlatitude mean westerly flow. Noteworthily, the extratropical response is in quadrature compared to 1xAMV; it is extremely similar in both ensembles not only in terms of spatial structure but also intensity, which seems to saturate with the amplitude of the AMV forcing. The alteration of the midlatitude North Atlantic dynamics in response to AMV can be interpreted as a combination of (i) local forcing associated with the subpolar gyre SST anomalies and (ii) tropical-extratropical teleconnectivity (Davini et al. 2015, Ruprich-Robert et al. 2017). The respective weight between the two mechanisms is expected to control the total model response and to explain part of its sensitivity to the amplitude of the AMV forcing as assessed here.

Regarding the tropical pathway of influence, evidence is provided in literature based on both theory and global circulation models that warmer SST in the subtropics is associated with increased precipitation on the northern flank of the climatological Inter Tropical Convergence Zone (ITCZ) yielding a Gill-Matsuno's type of atmospheric response in the Tropical Atlantic. Such a feature is consistently found in CNRM-CM5, which simulates, in all AMV ensembles, enhanced rainfall between the equator and 15ºN and a concomitant dipole in upper-tropospheric streamfunction straggling the equator, as depicted in Fig. 5. The anomalous anticyclonic circulations are located at 20º-30ºN and 10ºS and are maximum on the northwestern, respectively southwestern side of the main source of latent heat released throughout the troposphere by enhanced convection and ascendant motion (not shown), as featured in Fig. 4bed showing the vertical zonally averaged profile of specific humidity. The overall response is largely linear in the Tropics with respect to the amplitude of the AMV-forcing and spatially matches with the linear framework presented in Gill (1980, their Fig. 3) in presence of off-equatorial diabatic heating.

At midlatitudes, as a direct consequence of the anomalous anticyclonic circulation/enhanced momentum convergence due to the tropical wind divergence related to increased precipitation, there is an acceleration of the zonal wind at the entrance of the climatological upper-level subtropical jet around 30°N and 70°W. This acceleration extends downstream to the center of the North Atlantic towards the Azores in 2xAMV and 3xAMV (as featured from streamfunction anomalies in Fig. 5bc), whereas it is weak and rather confined to the western part of the basin in 1xAMV (Fig. 5a). This leads altogether in 2xAMV and 3xAMV to a strengthening of the westerly wind along the southern edge of the subtropical jet (as already described in Fig. 4fi from zonal averages), a feature which is not present in 1xAMV. At higher latitudes, circulation anomalies display a wave pattern along a southwest-northeast great circle from the Caribbean to Scandinavia in agreement with the classical stationary Rossby wave theory (Hoskins and Karoly, 1981). Again, signals are rather weak in 1xAMV (Fig. 5a) but are considerably reinforced in 2xAMV and 3xAMV with pronounced cyclonic circulation off Newfoundland and dominant anomalous anticyclonic flow from 55ºN northward (Fig. 5bc and Fig. 2kl); this is associated with a weakening of the westerly wind on the northern side of the climatological upper-level jet in the latter two ensembles (Fig. 4fi). Consistently, the AMV-forced response is then characterized by large-scale enhanced (reduced) baroclinicity in the southern (northern) side of the jet and equatorward shift in storm track driven by planetary wave changes that are reinforced and/or maintain through eddy-mean flow interaction and favored cyclonic Rossby wave breakings at short timescale (synoptic eddies, Rivière 2009, Davini and Cognazzo 2014), when the Atlantic is warmer (not shown). In 1xAMV, the alteration of the storm track is more regional and confined off Europe where a reduction occurs.

Despite differences in amplitude, these mechanisms share many features previously identified in Hodson et al. (2011), Peings et al. (2015), Davini et al. (2015) among others, in particular for the attribution of the midlatitude anomalous cyclonic circulation to a tropical forcing, which originates from the Caribbean basin in response to warmer Tropical Atlantic SST. For instance, there is a remarkable agreement between Fig. 5 and Fig. 2jkl in the present paper with Fig. 10 in Terray and Cassou (2002), based on an earlier version of the ARPEGE model used in sensitivity experiments to isolate the respective role of tropical versus extratropical North Atlantic SST anomalies, and with Fig. 9 in Ruprich-Robert et al. (2017). Results from decomposition of the daily circulation into weather regimes reveal a significant predominance of NAO- events at the expense of NAO+ in 2xAMV and 3xAMV (not shown), consistently with the mean circulation changes displayed in Fig. 2kl; this is indicative for a large contribution of the AMV tropical component in line with above-cited papers and Cassou et al. (2004). In 1xAMV, there is no changes in NAO-related regimes but a slight, albeit non-significant, prevalence of blocking circulations (not shown). We speculate here that the tropical influence is less dominant and that the extratropical SST component of the AMV is a key factor to explain the total response. Warmer SST in the subpolar gyre induces a reduction of the North Atlantic meridional temperature gradient exported upward throughout the troposphere with maximum loading at the intergyre around 45ºN; it is responsible for a collocated weakening of the westerly circulation as shown in Fig. 4c (Peings and Magnusdottir, 2014). Amplification due to sea ice loss is also expected to play a role in the slowdown of the jet (see Deser et al., 2015, Oudar et al., 2017 and the review of Screen et al., 2018 including CNRM-CM5). Note also that the anticyclonic circulation (Fig. 2j and Fig. 5a) is located between Iceland and the UK in 1xAMV, namely downstream to the maximum SST anomalies over the subpolar gyre (Fig. 1a) and related diabatic heating (precipitation anomalies in Fig. 5a), as opposed to 2xAMV and 3xAMV where the anomalous core in geopotential (Fig. 2kl) and upper-level streamfunction (Fig. 5bc) is centered over Easter Greenland. The 1xAMV pattern is consistent with an equivalent barotropic atmospheric response to extratropical SST anomalies resulting from changes in the position or strength of the stromtracks in presence of anomalous meridional SST gradient and related altered baroclinicity, as described in Kushnir et al. (2002).

Note finally that North Pacific-North Atlantic connection due to the remote effects of AMV on the Pacific basin-scale climate may also play a role. Consistently with Ruprich-Robert et al. (2017), results with CNRM-CM5 show a forced response in the Pacific that is reminiscent to the negative phase of the Pacific Decadal Variability (PDV) in SST. In terms of atmospheric circulation, slackened Aleutian Low as part of a larger scale Rossby wave pattern cascading into the North Atlantic, is also detectable in the model (not shown), but the amplitude of the Pacific-Atlantic teleconnection is lower in CNRM-CM5 compared to other models taken from the DCPP-C database (Ruprich-Robert et al. 2019). We thus interpret the dynamical response over Europe as primarily driven by the local Atlantic influence with some modulation from the Pacific-initiated wave train in line with Ding et al. (2017)'s findings from observations. The role of the PDV, which is likely function of AMV-forcing amplitude, however remains to be better quantified but a deeper analysis goes beyond the scope of the present paper.

*d. Mechanisms of the thermodynamical component of the AMV-forced response*

As above detailed, the thermodynamical component is computed as the difference between the total modeled response and the estimated contribution of the dynamical changes: it is a residual term encompassing multiple processes. Advection of heat at low-level atmosphere is one of them and has been shown to be a key factor to understand temperature anomalies over a given sector (see e.g. De Vries et al., 2013). It is assessed from the advection term $\mathbf{V}.\nabla T$, where $\mathbf{V}$ stands for wind speed and $\nabla T$ for temperature gradient usually taken at 850 hPa level to exclude turbulent and direct surface radiative influence in the boundary layer. In winter, the climatological westerly flow tends to advect relatively warm and humid oceanic air inland towards Europe. During positive AMV, an increase of the thermal advection by the climatological westerly flow is therefore expected because of warmer ocean but we showed that the latter thermodynamical term is counteracted by anomalous easterlies associated with the forced anticyclonic dynamical anomalies located off Europe (Fig. 3abc). As a final result, individual terms tend to cancel each other in 1xAMV, with yet slight weakening of the total advection along the Atlantic flank (except Scandinavia) thus contributing to cooling (Fig. 6a) and slight intensification around 55ºN in Central Europe leading to warming. Fig. 6bc shows a progressive reinforcement of the thermal advection with the amplitude of the AMV along the Atlantic flank. The larger changes are found in 3xAMV with an increase of the advection up to 30% over Germany contributing to large-scale warming from the UK to Poland. As the easterly wind anomalies are very similar in all AMV ensembles (Fig. 3), this suggests that the weight of the thermodynamical term in the total advection, a.k.a. the transport of temperature anomalies by the climatological westerly flow, becomes dominant and contributes to explain a significant fraction of the full T2m positive anomalies found in 2xAMV and 3xAMV (Fig. 2abc) in consistence with the outcomes of the thermodynamical-dynamical decomposition (Fig. 3a-f).

Changes in surface energy balance are also included in the temperature anomalies driven by thermodynamical processes. No significant anomalies of latent and sensible heat fluxes are observed over Europe (not shown). Significant changes in cloud cover (Fig. 7abc) are noted, with potential impacts on net longwave (Fig. 7def) and net shortwave (Fig. 7ghi) radiation at surface. In 1xAMV, a significant increase in longwave radiation and decrease in shortwave radiation are noted over the north of Germany, Poland etc. (Fig. 7dg), associated with an increase, although non significant, in cloud cover there (Fig. 7a). These anomalies are consistent with the radiative impact of clouds, with a greenhouse effect that tends to increase longwave radiation at surface and a parasol effect that tends to reduce shortwave radiation at surface. In 2xAMV, the cloud cover decreases almost everywhere over Europe, with significant values over the north of Poland again and over Greece and Turkey (Fig. 7b). These negative cloud anomalies are also associated with a significant increase in longwave radiation and a decrease in shortwave radiation at surface (Fig. 7eh). In 3xAMV, the cloud cover increase is particularly pronounced over the northeast of France, Benelux and the north of Germany (Fig. 7c), with a large and significant increase in longwave radiation there (Fig. 7f). The impact of cloud cover on net shortwave anomalies at surface is less clear (Fig. 7i). This might be related to the large decrease in snow cover over large parts of Europe seen in 3xAMV (Fig. 2i).

Changes in snow cover indeed impact the shortwave radiative budget at surface through the associated changes in surface albedo. The AMV-driven impact of albedo anomalies on shortwave radiation (Fig. 7jkl) is estimated thanks to the anomalous upward shortwave flux at surface $\Delta SW_\uparrow$ computed as follows:

$$\Delta SW_\uparrow = SW_{\downarrow AMV-} \cdot \Delta \alpha$$

where $SW_{\downarrow AMV-}$ is the downward shortwave radiation at surface in AMV- and $\Delta \alpha$ is the albedo anomaly between AMV+ and AMV- phases. Changes in snow cover are too weak in 1xAMV and 2xAMV to induce a significant upward shortwave radiation anomaly, except for a few points (Fig. 7jk). In 3xAMV, the snow cover anomalies are larger over Europe (Fig. 2l), and are associated with a large positive radiative response (Fig. 7l) with, for example, positive anomalies up to 1.5 W m$^{-2}$ obtained over Central Europe. It explains why the net shortwave radiation anomalies at surface are positive over the Alps, eastern Europe etc. despite the negative anomalies in downward shortwave radiation at surface associated with the increase in cloud cover (Fig. 7c).

As the increase in net longwave radiation at surface induced by an increase in cloud cover tends to be greater than the associated decrease in net shortwave radiation in winter over Europe, and additionally, given the decrease in upward solar radiation due to snow cover reduction, the net total radiation at surface tends to increase both in 1xAMV, 2xAMV and 3xAMV over most of Europe (Fig 7mno) although these differences are mostly significant in 3xAMV (Fig 7o).

The thermodynamical temperature anomalies seem to be mainly explained by three mechanisms: (i) the strengthening of the advection of warmer and moister oceanic air by the climatological westerly flow during positive AMV phases, (ii) the decrease of snow cover and (iii) the increase of clouds, both driving changes in shortwave and longwave radiation at surface. The increase in cloud cover may be related to the eastward advection of cloud and/or of warm and moist air by the mean flow from the Atlantic to the European cold continental areas where cloud formation would be enhanced.

## 4. Conclusions and discussion

In this study, the teleconnection between the AMV and the wintertime climate over Europe is assessed with the CNRM-CM5 coupled model via DCPP-C pacemaker experiments (Boer et al. 2016), in which the modeled North Atlantic SSTs are restored towards anomalies that are characteristic of the observed AMV. The sensitivity of the teleconnection to the AMV amplitude is evaluated thanks to three ensembles of simulations with different amplitudes of targeted SST anomalies. In the first ensemble (1xAMV), which strictly follows the DCPP-C coordinated protocol, the targeted SST anomalies correspond to one standard deviation on the observed AMV. They are respectively doubled and tripled for the 2xAMV and 3xAMV ensembles.

Fig. 8 wraps up our findings for surface temperature averaged over Europe. Positive AMV tends to be associated in winter with positive temperature anomalies especially in the 2xAMV and 3xAMV experiments as assessed from ensemble means; in 1xAMV, signals are very weak and barely significant (Fig. 8a). Spatial averages mask some regional features in 1xAMV with a slight cooling over a broad central Europe compensated by warming in Scandinavia and along the Mediterranean Sea to a lower extent. Precipitation anomalies tends to be positive over Europe, except Scandinavia, but their significance is marginal and does not evolve consistently with the amplitude of the AMV.

We apply a flow analog method in the three ensembles to decompose the total temperature and precipitation response in a dynamical part and a residual signal mostly including thermodynamical processes. During a positive phase of the AMV, in all the ensembles, the thermodynamical response is characterized by large-scale and positive T2m (Fig. 8b) and precipitation anomalies over most of Europe. Different mechanisms govern this net response: (i) the advection of positive temperature anomalies by the climatological westerly flow, (ii) the radiative effect of increase of cloud cover at the surface and (iii) the decrease of snow cover over Central Europe. The intensity of the thermodynamical warming migrates deeper and deeper inland from the Atlantic coast with the respect to the amplitude of the AMV-forcing with some positive feedback associated with the progressive snow cover disappearance (eastward retraction).

By contrast, the dynamical response is characterized by negative temperature (Fig. 8c) and precipitation anomalies mostly over the northern half of Europe, because of the presence of large-scale AMV-forced northeasterly wind anomalies that counter the climatological advection of relatively warm and moist air from the ocean. We speculate that the net response to AMV forcing in terms of atmospheric dynamics can be understood as a combined effect of extratropical and tropical influences: (i) the tropical branch of the AMV SST anomalies enhances local diabatic heating at the northern edge of the climatological ITCZ acting as a Rossby-wave source via a Gill-Matsuno's response, which cascades over Northern Europe; (ii) positive SST anomalies over the subpolar gyre and associated sea ice melting in all the Nordic Seas, responsible for polar amplification, lead to the development of an anomalous High at mid-to-high latitude. Preliminary results from additional twin ensemble experiments (also proposed in DCPP-C) where the 1xAMV full pattern is split into tropical and extratropical anomalies, tend to confirm our interpretation (not shown). We conclude that the tropical component dominates the total response in 2xAMV and 3xAMV with some modulation by the extratropical forcing whereas contributions are comparable in 1xAMV. This has some strong implication for impacts over Europe located at the tail end of the chain of influences. As shown in Fig. 3, despite large-scale easterly anomalies over most of the continent in all AMV ensembles, the precise position of the anomalous pressure centers of action in response to AMV matters a lot (especially for precipitation). The T2m and precipitation responses over Europe are found to be shifted northward as the AMV amplitude increases. Based on our interpretation, the position is very likely controlled by the respective weight between the strength, curvature, waviness and northeastward extension of the tropically-forced Rossby wave on one hand and the extratropical forcing included sea-ice on the other hand, with some possibility for partial non-additivity of the dynamical signals because of nonlinear processes (polar amplification, tropical convective responses etc.). We believe that the methodological framework proposed here, namely the decomposition of the AMV impacts over Europe into dynamical and thermodynamical components, could be a useful process-based approach to characterize and understand the inter-model differences regarding the AMV-forced teleconnection obtained from all the CMIP6 models involved in DCPP-C.

As a summary, the thermodynamical and dynamical impacts of the AMV on European temperatures tend to be opposed in CNRM-CM5 and confirm previously results shown by Yamamoto and Palter (2016) and O'Reilly et al. (2017) based on observations. For weak AMV forcing, both terms compensate each other and no significant impact of the AMV is obtained over Europe, while a significant warming is found during positive AMV phases in 2xAMV and 3xAMV experiments due to the thermodynamical response which becomes dominant. Noteworthily, the net temperature anomaly averaged over Europe scales linearly with the amplitudes of the AMV- SST anomalies mostly because of the thermodynamical component as evidenced from Fig. 8. But recall that is not the case regionally, therefore challenging the validity of so-called pattern-scaling technique to evaluate teleconnectivity and related impacts associated with AMV-type of variability. Dividing respectively by two and three the 2xAMV and 3xAMV forced response in order to get a proxy for the 1xAMV spatial fingerprint, does not reproduce the actual map of the 1xAMV response obtained from the model. The pattern correlation in T2m between the patterns from the 2xAMV and 3xAMV ensembles and the real 1xAMV outcome is equal to 0.83 and 0.67, respectively. Same conclusions are found for precipitation with values equal to 0.47 and 0.33. The limit for pattern scaling is speculated here to be related to the relative changes between dynamical and thermodynamical influences partly governed by nonlinear processes (polar amplification, eddy-mean flow interaction which is crucial at the tail end of the cascading Rossby waves over Europe, snow cover effects, etc.).

Assessing the true degree of linearity of the response is further complicated by some intrinsic limitations related to the experimental protocol based on pacemaker techniques. Since the SST restoring coefficient is fixed, it is more efficient in the tropics than in the extratropics because its strength is function of the ocean mixed layer depth. The actual SSTs are therefore closer to the targeted SSTs in the tropical band dominated by pronounced stratification (shallow mixed layer) than at midlatitudes characterized by deep mixing (thick mixed layer) (not shown). More weight might be therefore artificially given to the AMV-forcing originating from the tropics relative to the one induced by extratropical SST anomalies. The relative weight between tropical and extratropical influences is not conserved when the AMV SST anomalies are multiplied by 2 and 3 as done in our study; this hampers a clean investigation of the linearity. We performed additional test experiments with a restoring coefficient that varies with the mixed layer depth as in Ortega (2017) but only for 1xAMV. Preliminary results suggest that the model response presented here for 1xAMV is robust (not shown) but further analyses would be needed to firmly conclude.

Fig. 8 also highlights the weak signal-to-noise ratio of the AMV-forced temperature response over Europe. Even if the spread in North Atlantic SST anomalies for a given AMV forcing is very constrained by the restoring framework, a very large inter-member spread is noted in all ensembles for the mean temperature response; it is maximum for the thermodynamical term. The spread is so large that even in the 3xAMV (2xAMV) experiment, 7 (16) members out of 40 have their 10-yr averaged temperature response over Europe of negative sign, despite overall warming effect of the AMV. Note that the AMV experiments are named according to the targeted SSTs and not to the actual SST anomalies obtained in the pacemaker ensembles. As shown in Fig. 1, the actual SST values in 2xAMV and 3xAMV are closer to the targeted SST of 1xAMV and 2xAMV, respectively. As a result, the actual SST anomalies in 2xAMV are far from rare, while those in 3xAMV becomes extreme but may still be observed since they correspond to two standard deviation of the observed AMV over the historical period.

A last perspective but with potentially strong implication would be to evaluate the sensitivity of the AMV-forced teleconnectivity to the model mean background state. In this work, we investigated the impact of the AMV in so-called pre-industrial climate. Nothing guarantees that the AMV teleconnection over Europe is independent of the climate mean state and could then change or has already changed as climate is warming due to anthropogenic factors. New twin experiments, in which the North Atlantic SST would be restored to the same anomalies as in this study but with a mean state characteristic of current climate (about +1ºC) or future climate warming targets depending on future CO2 emission scenarios could be of particular interest. Beyond climate change, linking AMV-forced response to model mean states (and consequently intrinsic biases) could be also a pertinent framework to understand the CMIP6 models diversity within the coordinated DCPP initiative.

**Acknowledgments:**

This work was supported by a grant from Electricité de France (EDF) and by the French National Research Agency (ANR) in the framework of the MORDICUS project (ANR-13-SENV-0002). The authors are grateful to Marie-Pierre Moine, Laure Coquart and Isabelle d'Ast for technical help to run the model. Computer resources have been provided by Cerfacs. The Figures were produced with the NCAR Command Language Software (http://dx.doi.org/10.5065/D6WD3XH5).

Table 1: Mean RMSE and spatial correlation between the analog daily SLP and the actual daily SLP estimated from all winter (Dec. 1st-Feb 28th) days over the 40 members and 10 years in 1xAMV+, as a function of the number of analogs and domains.

| Number of analogs | 1 | 5 | 10 | 15 | 20 | 30 |
|---|---|---|---|---|---|---|
| Europe (35°N-70°N, 15°W-35°E) | | | | | | |
| RMSE (hPa) | 3.9 | 2.8 | 2.7 | 2.8 | 2.8 | 2.8 |
| Correlation | 0.92 | 0.96 | 0.96 | 0.96 | 0.96 | 0.96 |
| North Atlantic – Europe (20°N-80°N, 80°W-30°E) | | | | | | |
| RMSE (hPa) | 3.8 | 4.3 | 4.1 | 4.1 | 4.1 | 4.1 |
| Correlation | 0.81 | 0.90 | 0.92 | 0.92 | 0.92 | 0.93 |

Table 2: Same as Table 1, but for T2m.

| Number of analogs | 1 | 5 | 10 | 15 | 20 | 30 |
|---|---|---|---|---|---|---|
| Europe (35°N-70°N, 15°W-35°E) | | | | | | |
| RMSE (°C) | 3.7 | 2.8 | 2.7 | 2.7 | 2.7 | 2.7 |
| Correlation | 0.87 | 0.92 | 0.92 | 0.93 | 0.93 | 0.93 |
| North Atlantic – Europe (20°N-80°N, 80°W-30°E) | | | | | | |
| RMSE (°C) | 3.8 | 2.9 | 2.8 | 2.7 | 2.7 | 2.7 |
| Correlation | 0.85 | 0.91 | 0.92 | 0.92 | 0.92 | 0.92 |

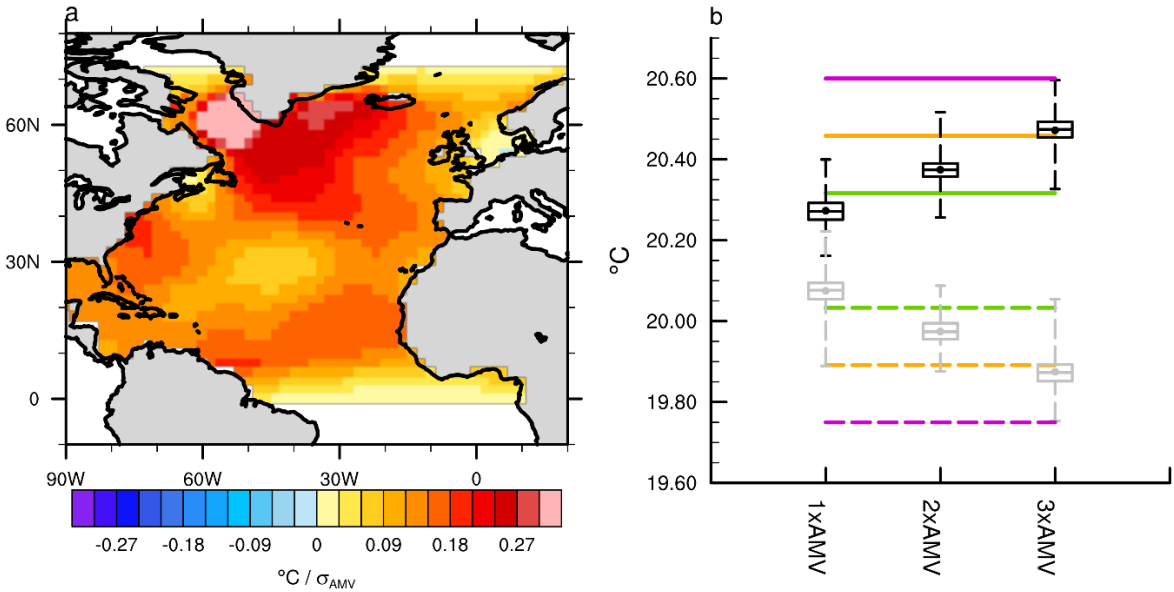

Fig. 1: (a) Anomalous SST pattern used for restoring and taken from input4MIPs archive (units are ºC/σ(AMV), shading interval is every 0.03ºC). (b) Simulated raw annual SST averaged over the North Atlantic restored sector for AMV+ (black) and AMV- (gray) experiments. Each boxplot stands for the distribution obtained from 360 years for each ensemble (40-members × 9-years, the first year being discarded to account for the model SST initial adjustment to restoring). The top (bottom) of the box represents the first (last) tercile of the distribution and the upper (lower) whisker represents the first (ninth) decile. Dots and inside-line stand for the mean and the median of the distribution, respectively. The green, orange and magenta horizontal lines show the SST targets for the 1xAMV, 2xAMV and 3xAMV ensembles corresponding to 1, 2 and 3 standard deviations of the observed AMV index, respectively. Solid and dashed stands respectively for AMV+ and AMV- experiments.

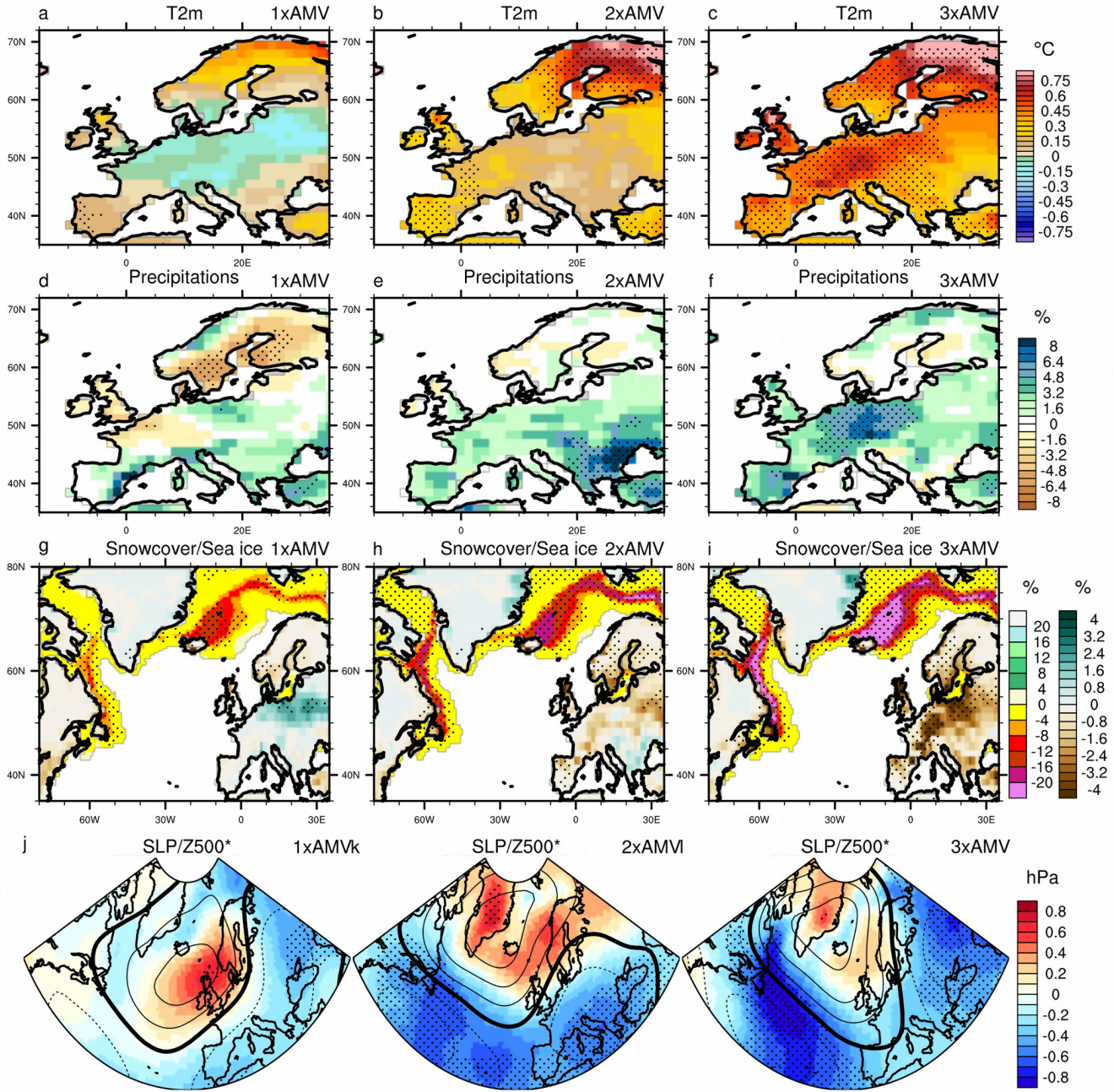

Fig. 2: AMV-forced anomalies for December-February seasonal mean for T2m (abc, shading interval is 0.075°C), precipitation (def, in relative percentage; shading interval is 0.8%), land snow cover and sea-ice (ghi, shading intervals are 0.4% and 4%, respectively) and Z500* (contour interval is 4 m and the thicker black contour stands for the zero line) superimposed on MSLP (shading interval is 0.1hPa) for 1xAMV (left), 2xAMV (center) and 3xAMV (right). Stippling indicates regions that are above the 95% confidence level of statistical significance based on two-sided Student's t-test.

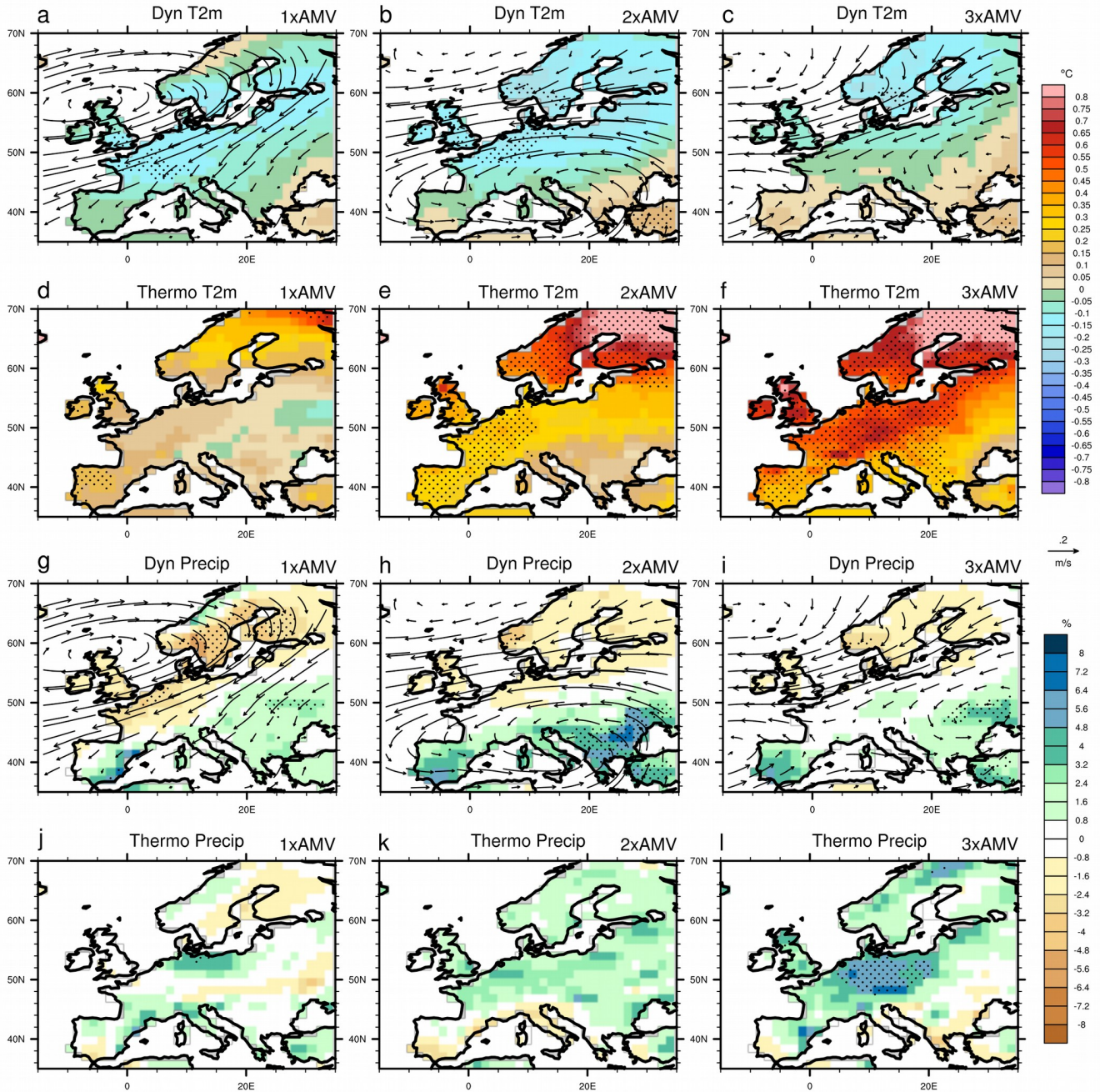

Fig. 3: Decomposition of the AMV-forced anomalies into dynamical and thermodynamical components estimated from analog reconstruction for December-February seasonal mean for T2m (first and second rows respectively, shading interval is 0.05 °C) and for precipitation (third and fourth rows respectively, in relative percentage; shading interval is 0.8%). See section 2.b for the description of the method. Stippling indicates regions that are above the 95% confidence level of statistical significance based on two-sided Student's t-test.

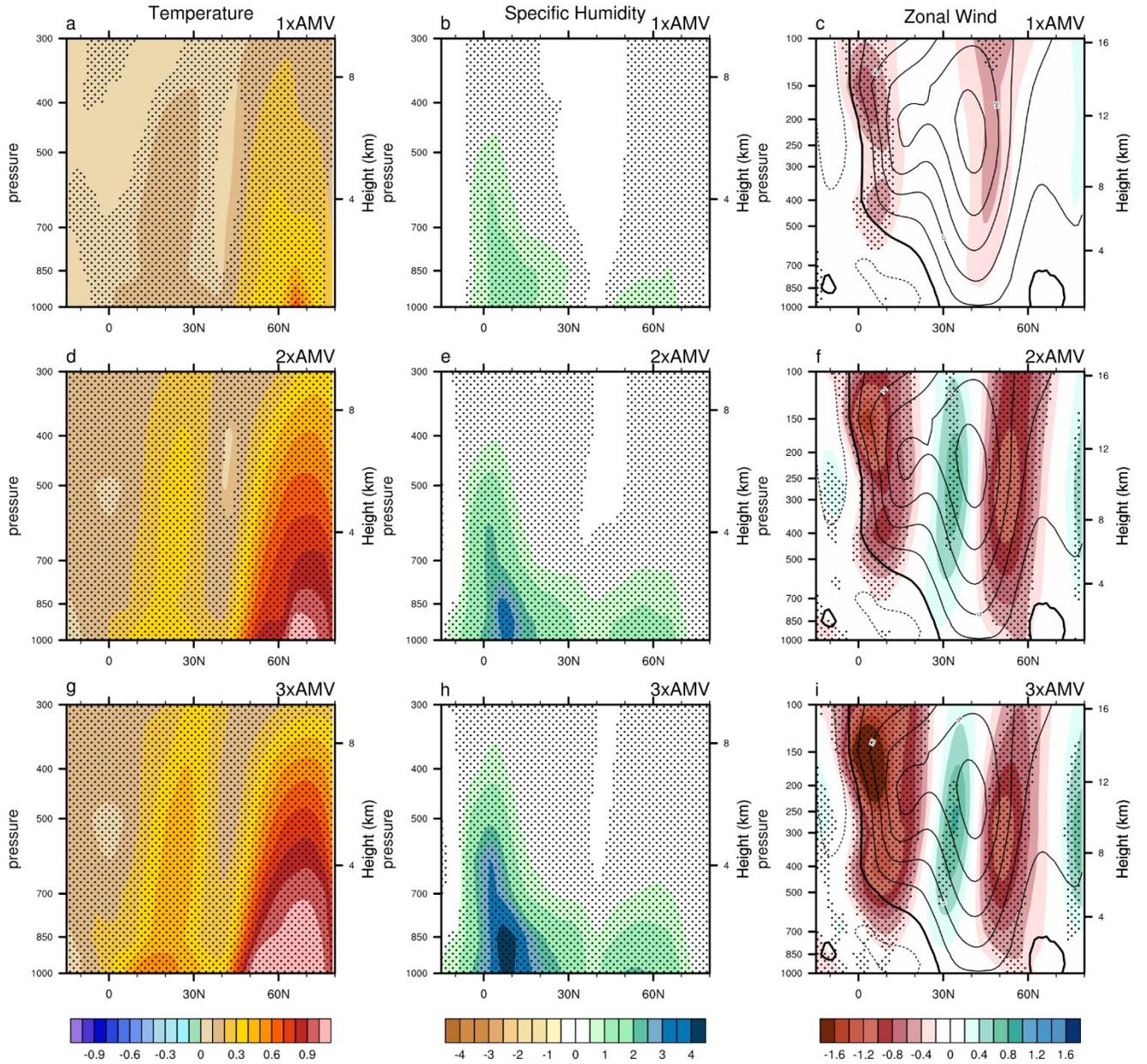

Fig. 4: Zonal average over the North Atlantic (15°S-80°N, 80°W-15°W) of the AMV-forced anomalies for December-February seasonal mean for temperature (first column, shading interval is 0.1°C), specific humidity (second column, in $10^{-4}$ kg kg$^{-1}$; shading interval is $0.5 \cdot 10^{-4}$ kg kg$^{-1}$), zonal wind (third column, shading interval is 0.2 m s$^{-1}$) with the climatological value superimposed from AMV- (contour interval is 4 m s$^{-1}$ and the thicker black contour stands for the zero line) for 1xAMV (abc), 2xAMV (def) and 3xAMV (ghi). Stippling indicates regions that are above the 95% confidence level of statistical significance based on two-sided Student's t-test.

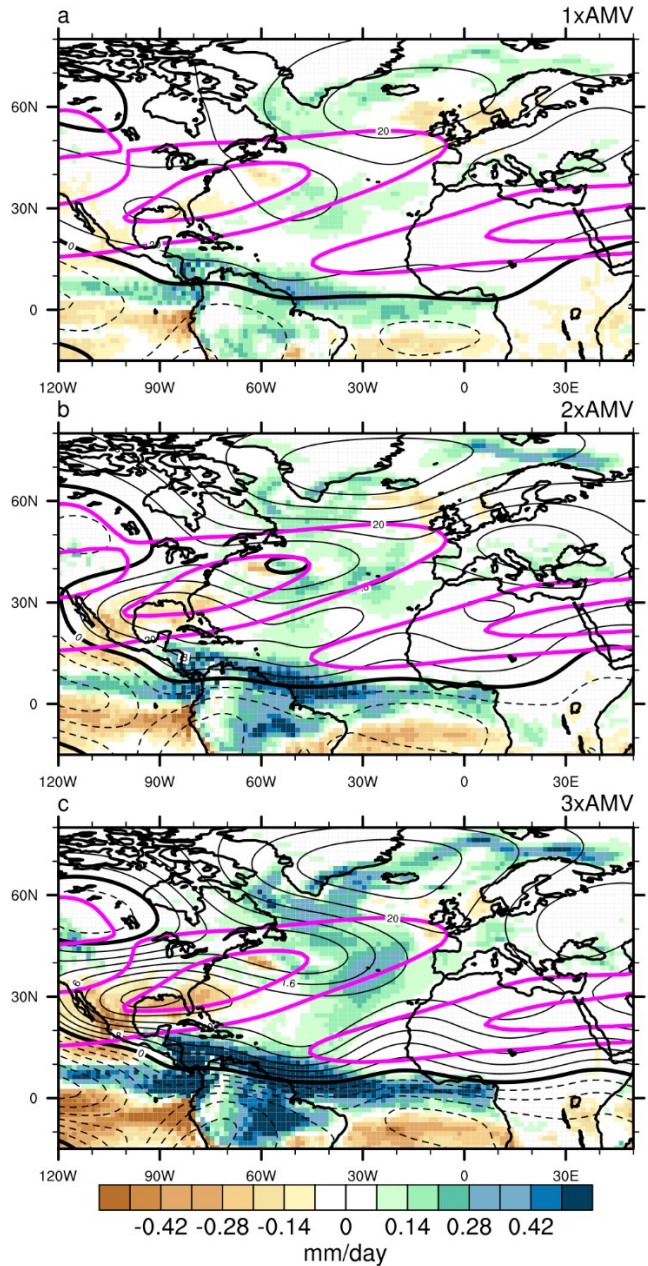

Fig. 5: AMV-forced anomalies for December-February seasonal mean for precipitation (shading interval is 0.07 mm day$^{-1}$), stream function at 200 hPa (black contour interval is 0.2 x 10$^6$ m² s$^{-1}$, the thicker black contour stands for the zero line) for 1xAMV (a), 2xAMV (b) and 3xAMV (c). Climatological zonal wind speed at 300 hPa from AMV- is superimposed (2 magenta contours at 20 and 25 m s$^{-1}$).

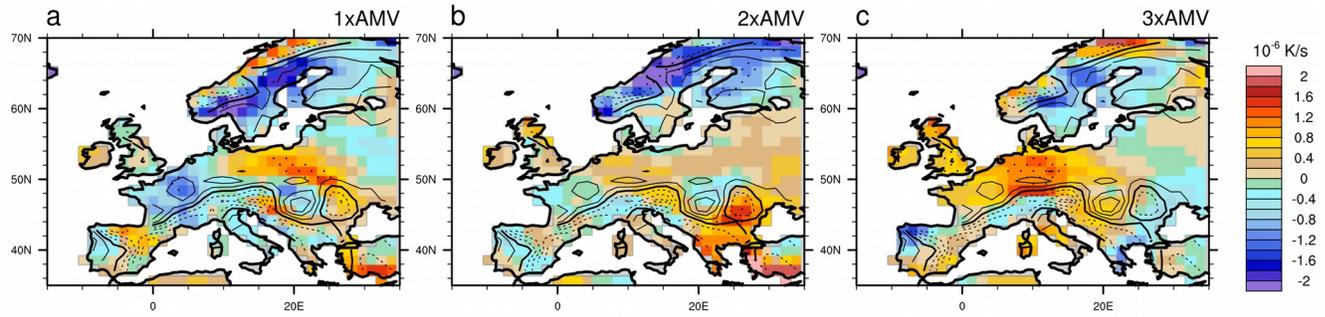

Fig. 6: AMV-forced anomalies for December-February seasonal mean for temperature advection anomalies at 850 hPa (shading interval is 0.2 $10^{-6}$ K s$^{-1}$) for 1xAMV (a), 2xAMV (b) and 3xAMV (c). Climatological advection from AMV- is superimposed (black contour interval is $10^{-6}$ K s$^{-1}$, the thicker black contour stands for the zero line). Stippling indicates regions that are above the 95% confidence level of statistical significance based on two-sided Student's t-test.

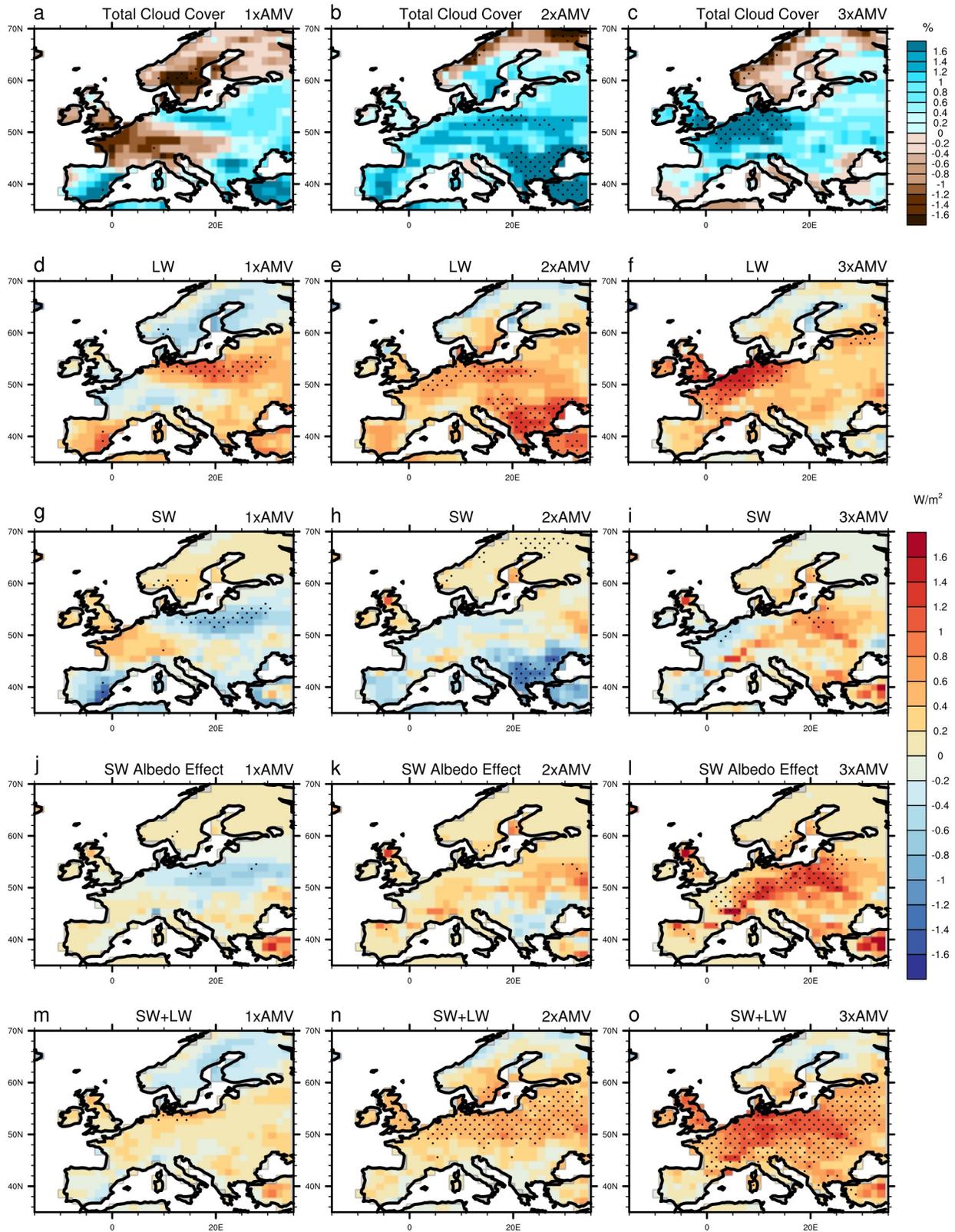

Fig. 7: AMV-forced anomalies for December-February seasonal mean for total cloud cover (abc, shading interval is 0.2%), net longwave radiation at surface (def, shading interval is 0.2 W m$^{-2}$), net shortwave radiation at surface (ghi), radiative effect due to surface albedo changes in shortwave (jkl) and net shortwave and longwave radiation at surface (mno) for 1xAMV (left), 2xAMV (center) and 3xAMV (right). Positive values represent energy moving towards the surface. Stippling indicates regions that are above the 95% confidence level of statistical significance based on two-sided Student's t-test.

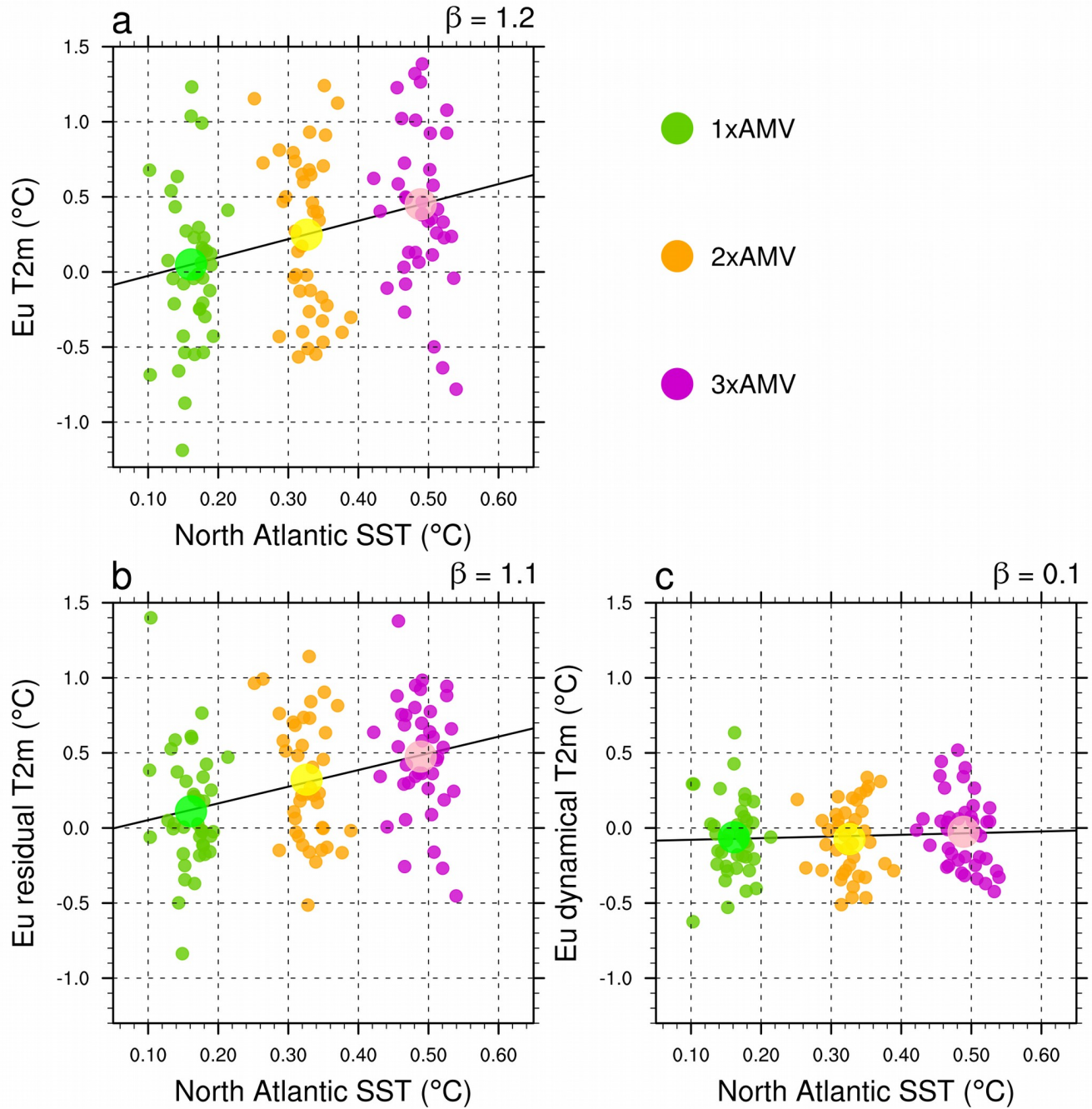

Fig. 8: Spatial average of AMV-forced anomalies for December-February seasonal mean of T2m anomalies over Europe (same domain as in Fig. 2) versus North Atlantic SST (0°-60°N) for the dynamical part (a), thermodynamical (or residual) part (b) and total anomalies (c) for 1xAMV (green), 2xAMV (orange) and 3xAMV (magenta). The small dots represent the 10-yr mean response of each member and the big dot stands for the ensemble mean. The slope $\beta$ obtained from the linear regression between the T2m and the SST anomalies distributions from all the experiments is given in the upper right title of each panel.